\documentclass{article}
\usepackage{spconf}
\usepackage{amssymb}
\usepackage{amsmath}
\usepackage{amsthm}
\usepackage{graphicx}
\usepackage{algorithm}
\usepackage{algorithmic}
\usepackage{cite}
\usepackage{bm}
\usepackage{float}
\usepackage{multirow}
\usepackage{color}
\usepackage{subfigure}
\usepackage{caption}
\usepackage{epstopdf}
\usepackage{hyperref}
\usepackage{pgfplots}
\usepackage{pgfplotstable}

\title{ECONOMICS OF SEMANTIC COMMUNICATION SYSTEM IN WIRELESS POWERED INTERNET OF THINGS}
\name{Zi Qin Liew$^1$, Yanyu Cheng$^1$, Wei Yang Bryan Lim$^1$, Dusit Niyato$^2$, Chunyan Miao$^{1,2}$, Sumei Sun$^{3}$}
\address{$^1$Alibaba-NTU Singapore Joint Research Institute, Nanyang Technological Univerity, Singapore \\ $^2$School of Computer Science and Engineering, Nanyang Technological University, Singapore \\ $^3$Institute of Infocomm Research, Agency for Science, Technology and Research, Singapore}
\usepackage[font=small]{caption}

\begin{document}
%
\maketitle

\begin{abstract}
The semantic communication system enables wireless devices to communicate effectively with the semantic meaning of the data. Wireless powered Internet of Things (IoT) that adopts the semantic communication system relies on harvested energy to transmit semantic information. However, the issue of energy constraint in the semantic communication system is not well studied. In this paper, we propose a semantic-based energy valuation and take an economic approach to solve the energy allocation problem as an incentive mechanism design. In our model, IoT devices (bidders) place their bids for the energy and power transmitter (auctioneer) decides the winner and payment by using deep learning based optimal auction. Results show that the revenue of wireless power transmitter is maximized while satisfying Individual Rationality (IR) and Incentive Compatibility (IC).
\end{abstract}
\section{Introduction}
\label{sec:intro}
Internet of Things (IoT) enables and provides many smart systems and intelligent services such as smart home, smart cities and smart healthcare. The success of IoT heavily relies on communication systems that can accurately and effectively transmit information from transmitter to receiver. Recent advancements in natural language processing (NLP) \cite{vaswani2017attention}, \cite{devlin-etal-2019-bert} and Deep Learning (DL) enabled end-to-end (E2E) communication system \cite{guler2018semantic} have enabled semantic level communication systems for text transmission. While conventional communications focus on reducing the bit-error rate (BER) or symbol-error rate (SER) \cite{al2012efficient}, \cite{mirahmadi2013ber}, semantic communication aims to transmit the information relevant to the transmission goal. Recent works showed that the semantic communication system is robust at low signal-to-noise (SNR) region \cite{xie2021deep}, \cite{xie2020lite}, which helps to increase the reliability of IoT services. The ability to communicate effectively with the semantic meaning behind digital bits can inspire more smart applications of IoT. For example, in text transmission, semantic communication system can process text in the semantic domain by extracting the meanings of the text, filtering out the irrelevant information. For practical implementation in the IoT devices, the size and complexity of the semantic communication model can be reduced by model compression as verified in \cite{xie2020lite}.

Other than the efficiency of a communication system, data transmission of IoT is also limited by the amount of energy on IoT devices. One of the energy sources of IoT devices is energy harvested from a hybrid access point (H-AP) \cite{ramezani2017toward},\cite{chae2018simultaneous}. In such wireless powered network, IoT devices use the harvested energy to transmit data back to H-AP. However, the rules of energy allocation from H-AP to the IoT devices in the semantic communication system is not well studied.

\begin{figure}[htb]
  \centering
  \centerline{\includegraphics[width=5cm]{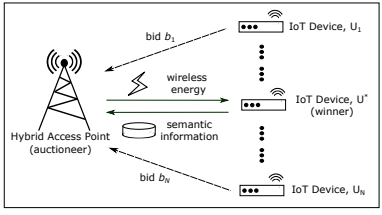}}
\caption{IoT devices bid for the energy from Hybrid Access Point}
\label{fig:iotbid}
\end{figure}

In this paper, we consider the energy allocation problem as an incentive mechanism design problem where the winner and payment of the energy are decided by a DL based auction mechanism \cite{dutting2019optimal}. As compared to traditional auction, DL based auction can maximize the revenue of seller while keeping the properties of individual rationality and incentive compatibility. In our system model, IoT devices (bidders) bid for the wireless energy from the H-AP (auctioneer) to transmit the encoded semantic information as shown in Fig. \ref{fig:iotbid}. The main contributions of our paper are (1) We propose an energy allocation framework to support wireless powered IoT that adopts semantic communication system. Different from previous works \cite{wu2015energy}, \cite{chingoska2016resource} on the resource allocation of wireless powered devices in conventional communication networks, our model tackles the energy constraint with an economic approach by designing an incentive mechanism for the energy allocation framework. (2) We propose an energy valuation procedure based on the sentence similarity score and bilingual evaluation understudy (BLEU) score \cite{papineni2002bleu} of the semantic communication system. While many recent works have focused on improving the performance of semantic communication system \cite{xie2020lite},\cite{xie2021deep},\cite{farsad2018deep}, few works have addressed the energy constraint in a wireless powered semantic communication system. Our model provides an incentive mechanism to derive the valuation of the energy. (3) We use DL based auction mechanism to decide the allocation of energy. In contrast to traditional auctions, the DL based auction maximizes the revenue of H-AP (auctioneer), without losing the properties of an optimal auction, i.e., IR and IC.

\section{System Model, Problem Formulation, and Methodology}
\label{sec:sysmodel}
\subsection{Wireless Powered Communication Network (WPCN)}
\label{ssec:wpcn}
\sloppy We consider a WPCN with a $K$-antenna H-AP and $N$ single-antenna users $\{\mathrm{U}_1, \mathrm{U}_2,\ldots, \mathrm{U}_N\}$, where the harvest-then-transmit protocol \cite{ju2013throughput} is adopted.
At a specific time, only one user can be served, and the beamforming vector at the H-AP is designed for this user.
The user first harvests the energy from the beam over a duration $\tau$, and then transmits the information to the H-AP over a duration $\tau_n'$. 
Here, the information-transmission time for the different users is different, and its value is introduced in the following.

Let $\mathbf{h}_n$ and $\mathbf{g}_n$ denote the $K$-dimensional downlink and uplink channel vectors between the H-AP and $\mathrm{U}_n$, respectively.
Each entry comprises two components, i.e., path loss and small-scale fading, where the small-scale fading is assumed to follow an independent and identically distributed (i.i.d.) Rayleigh fading model.
Following \cite{huang2015performance}, we consider that the channel state information (CSI) can be perfectly estimated by the H-AP, and the CSI remains constant in each period.

For the maximum ratio transmission of downlink energy transfer, the optimal beamforming vector for $\mathrm{U}_n$ is designed as $\mathbf{w}_n^*=\mathbf{h}_n/\lVert\mathbf{h}_n\rVert$ \cite{huang2015performance}.
The energy harvested from the noise and other users is negligible, and the harvested energy of $\mathrm{U}_n$ is given by $E_n=\eta \tau P \lVert\mathbf{h}_n\rVert^2$ , where $P$ is the transmit power of the H-AP, and $\eta \in[0,1]$ is the energy conversion efficiency.

After harvesting energy, $\mathrm{U}_n$ transmits the information to the H-AP, and the observed signal at the H-AP is expressed as $\mathbf{y}_n=\sqrt{\frac{E_n-E_{cir}}{\tau_n'}} \mathbf{g}_n x_n + \mathbf{n}$, where $E_{cir}$ is the energy consumption from the circuit, $x_n$ is $\mathrm{U}_n$'s transmitted signal, and $\mathbf{n}$ is the additive Gaussian noise with zero mean and variance matrix $\sigma^2 \mathbf{I}_N$.
We adopt the maximum ratio combining (MRC) technique at the AP to maximize the received SNR \cite{huang2015performance}.
Thus, the SNR at the AP is given by
\begin{equation}
\mathbf{\gamma}_n=\frac{(E_n-E_{cir})\lVert\mathbf{g}_n\rVert^2}{\tau_n' \sigma^2}
=\frac{\tau\rho \lVert\mathbf{h}_n\rVert^2 \lVert\mathbf{g}_n\rVert^2- \xi\lVert\mathbf{g}_n\rVert^2}{\tau_n'},
\end{equation}
where $\rho = \eta P/\sigma^2$, and $\xi=E_{cir}/\sigma^2$.

We assume that the fixed-rate transmission is adopted in this paper. When the SNR is greater than a threshold $\phi$, the message is transmitted with a fixed data rate, which is $R = \log_2(1+ \phi)$.
To best use the resource, for $\mathrm{U}_n$, its SNR should equal $\phi$, so that the information-transmission time is given by
\begin{equation}
\tau_n'^* = \frac{\tau\rho \lVert\mathbf{h}_n\rVert^2 \lVert\mathbf{g}_n\rVert^2- \xi\lVert\mathbf{g}_n\rVert^2}{\phi}.
\end{equation}
Accordingly, the number of bits that $\mathrm{U}_n$ can transmit by using the harvested energy is given by
\begin{equation}
\mathbb{N}_n = R \tau_n'^*
=\log_2(1+ \phi) \frac{\tau\rho \lVert\mathbf{h}_n\rVert^2 \lVert\mathbf{g}_n\rVert^2- \xi\lVert\mathbf{g}_n\rVert^2}{\phi}.
\label{numbit}
\end{equation}

\subsection{DL Enabled Semantic Communication Systems}
\label{deepsc}

We consider the $N$ single-antenna users as $N$ IoT devices that perform text data transmission with DL enabled semantic communication systems, e.g., voice controlled smart devices (Google Nest Hub, Amazon Echo, Apple HomePod). In DL enabled semantic communication system, collected sentences, $\mathbf{S}=[s_1,s_2,\ldots,s_{N_s}]$, are encoded by semantic encoder and channel encoder. The encoded signal can be represented by $\mathbf{X} = enc_c(enc_s(\mathbf{S})) \label{enc}$, where $\mathbf{X} \in \mathbb{R}^{N_s \times L \times D}$, $N_s$ is the number of sentences, $L$ is the sentence length, $D$ is the output dimension of channel encoder, $enc_c(\cdot)$ is the channel encoder, and $enc_s(\cdot)$ is the semantic encoder. Note that all input is padded to length $L$ before passing to the encoders. After energy harvesting, IoT devices transmit encoded information to H-AP. At the receiver's end, signal received can be expressed as $\mathbf{Y} = \mathbf{HX}+\mathbf{A}$, where $\mathbf{H}$ is the channel gain between the transmitter and receiver and $\mathbf{A} \sim \mathcal{N}(0,\sigma^2_n)$ is the additive white Gaussian noise (AWGN). The decoded sentences are given by $\widehat{\mathbf{S}} = dec_s(dec_c(\mathbf{Y}))$,  where $dec_s$ and $dec_c$ are the semantic decoder and channel decoder of the receiver.

In this paper, we adopt the network architecture of DeepSC \cite{xie2021deep} where semantic encoder and decoder are implemented as multiple Transformer \cite{vaswani2017attention} encode and decode layers, and channel encoder as dense layers with different units. There are also other network architectures used for semantic encoder and decoder, e.g., bidirectional long short term memory (BLSTM) network in \cite{farsad2018deep}. Our incentive mechanism can be easily extended to other network architectures by following the same evaluation procedure.

BLEU score and sentence similarity are two of the critical performance metrics of the semantic communication system. BLEU score measures the exact matching of words in the original and recovered sentences without considering their semantic information. In contrast to BLEU score, sentence similarity is calculated by the cosine similarity of the extracted semantic features from original and recovered sentences. In our model, a pre-trained Bidirectional Encoder Representations from Transformers (BERT) \cite{devlin-etal-2019-bert} model is used for semantic features extraction. Let $\mathbf{s}$ and $\widehat{\mathbf{s}}$ denote one sentence from $\mathbf{S}$ and $\widehat{\mathbf{S}}$, respectively. BLEU score can be expessed as $\log \text{BLEU} = \min \left(1-\frac{l_{\widehat{\mathbf{s}}}}{l_{\mathbf{s}}},0\right) + \displaystyle\sum_{i=1}^{I} u_i \log p_i$, where $l_{\mathbf{s}}$ and $l_{\widehat{\mathbf{s}}}$ are the length of the original and recovered sentences respectively, $u_i$ is the weight of $i$-grams, and $p_i$ is the $i$-grams score, which is given by $p_i = \frac{\sum_k \min(C_k (\widehat{\mathbf{s}}), C_k(\mathbf{s}))}{\sum_k \min(C_k(\widehat{\mathbf{s}}))}$, where $C_k (\cdot)$ is the frequency count function for the $k$-th element in $i$-th grams.
The sentence similarity is given by $similarity(\widehat{\mathbf{s}},\mathbf{s}) = \frac{\mathbf{B}(\mathbf{s}) \cdot \mathbf{B}(\widehat{\mathbf{s}})^T}{\lVert\mathbf{B}(\mathbf{s})\rVert \lVert\mathbf{B}(\widehat{\mathbf{s}})\rVert}$ , where $\mathbf{B}(\cdot)$ is a pre-trained BERT model. 

In general, the BLEU score and similarity score are higher when encoded information has more features. As the output dimension, $D$ increases, the feature size of the encoded sentence increases, the decoder has more information to recover better the original sentence, and achieves higher similarity between the original and recovered sentence. However, increasing $D$ comes at the cost of data size, and the amount of data that IoT devices can send is limited by $\mathbb{N}_n$ derived from Equation \eqref{numbit}. Specifically,  the BLEU score and similarity score of device $\mathrm{U}_n$ can be expressed as 
$s_n = f_{sim}(D) = f_{sim}(\frac{\mathbb{N}_n}{N_s\times L \times b_f}) $, and $BLEU_n = f_{BLEU}(D) = f_{BLEU}(\frac{\mathbb{N}_n}{N_s\times L \times b_f}) $, respectively, where  $f_{sim}(\cdot)$ and $f_{BLEU}(\cdot)$ are simple lookup to obtain the scores of the model, and $b_f$ is the number of bits used by a unit feature. A unit feature is a single entry of $\mathbf{X} \in \mathbb{R}^{N_s \times L \times D}$, and $b_f$ is the number of bits used to represent a float type data.

We consider that the model on IoT device is pre-trained with $D=d_1$. Under the limited data budget, $\mathbf{X} \in \mathbb{R}^{N_s \times L \times d_1}$ is trimmed to $\mathbf{X}' \in \mathbb{R}^{N_s \times L \times d_2}$, where $0<d_2<d_1$. At the receiver, the trimmed data is recovered by sampling values from $\mathcal{N}(\mu_{X'},\,\sigma^{2}_{X'})$, where $\mu_{X'}$ and $\sigma^{2}_{X'}$ are the mean and variance of the entries of $\mathbf{X}'$. To obtain $f_{sim}(\cdot)$ and $f_{BLEU}(\cdot)$, we first train the DeepSC model with $D=d_1$ and use the trained model to evaluate the similarity scores when $D = d$, $\forall d \in [1,d_1]$. Then, we can obtain $f_{sim}(\cdot)$ and $f_{BLEU}(\cdot)$ from the evaluation results of test datasets.

\subsection{DL Based Auction}
\label{deepauction}
 As H-AP can serve only one user at a specific time, the IoT devices need to compete for the energy from H-AP. Given the energy constraint, we propose an economic approach where IoT devices (bidders) bid for the wireless energy from the H-AP (auctioneer) in a single-item auction. IoT devices will bid for the energy according to the similarity score and BLEU score derived from Section \ref{deepsc}. Generally, when the BLEU score and similarity score are higher, the device has more incentive to pay a higher price for the energy. The valuation of energy can be expressed as $v_n = j_ns_n + m_nBLEU_n$, where $j_n$ and $m_n$ are the preferences of the device in similarity score and BLEU score respectively, and $j_n+m_n=1$. If $j_n < m_n$, it indicates that the device has more interest in the exact recovery of words whereas $j_n > m_n$ indicates higher interest in the matching of semantic meaning. For example, some medical devices would have higher $m_n$ because the exact recovery of medical terms is more important, whereas devices that collect data for text classification would have higher $j_n$. 

In every round of single-item auction, H-AP, i.e., the auctioneer collects the bids $(b_1,b_2,\ldots,b_N)$ from all IoT devices, i.e., bidders, and then decides the winner, $\mathrm{U}^*$, and corresponding payment price, $\theta_{\mathrm{U}^*}$. The utility of the device is given by $u_n = b_n - \theta_{\mathrm{U}^*}$, if the device is the winner and $u_n = 0$ otherwise. Traditional single-item auctions such as the first-price auction and Second-Price Auction (SPA) can be used to determine the winner and price. For an auction to be optimal \cite{myerson1981optimal}, it should attain the properties of IR and IC. IR guarantees that the IoT devices have non-negative utility by participating in the auction, i.e., $u_n \geq 0$. IC ensures that each device will submit bids according to their true valuations, i.e., $b_n = v_n$, regardless of the actions of other devices, and the utility of each device is maximized by submitting the truthful bid. In the first-price auction, the highest bidder wins and pays the exact bid submitted, maximizing the revenue gain of H-AP but does not guarantee IC. In SPA, the highest bidder wins but pays the price of second highest bidder. SPA ensures IC but does not maximize the revenue of H-AP. 

To determine the energy allocation of H-AP, we adopt a DL based optimal auction mechanism \cite{dutting2019optimal} that can maximize the revenue of seller while keeping the properties of IC and IR. The auctioneer (H-AP) does not have a priori knowledge about the bidders and optimal decisions in determining the winner. Nevertheless, the H-AP can learn from experience and adjust the auction decision using DL based optimal auction. Due to space constraints, the details of the deep learning network used in \cite{dutting2019optimal} is omitted. The algorithm of the deep learning network is shown in Algorithm \ref{alg:aucalg}. We denote the monotone transform function as $\Phi$, allocation rule as $z_n$, conditional payment rule as $\theta_n$, number of groups for linear function as $Q$, linear functions in each group as $S$, weights and bias parameters as $\mathbf{w}$ and $\bm{\beta}$ respectively, and the expected revenue of H-AP as loss function, $\hat{R}(\mathbf{w} ,\bm{\beta})$.

\begin{algorithm}
  \caption{DL Based Auction}\label{alg:aucalg}
  \hspace*{\algorithmicindent} \textbf{Input:} Bids of IoT devices $b_i=(b_1^i,b_2^i,\ldots, b_N^i)$\\
  \hspace*{\algorithmicindent} \textbf{Output:} Revenue gain by H-AP
  \begin{algorithmic}[1]
    \STATE \textbf{Initialization:} $\mathbf{w} \in \mathbb{R}_{+}^{I\times QS}, \bm{\beta} \in \mathbb{R}^{I\times QS} $
    \WHILE{Loss function $\hat{R}(\mathbf{w,\bm{\beta}})$ is not minimized}
    \STATE Compute transformed bids $\bar b_n^i = \theta_n(b_n^i) = \min_{q\in Q}\max_{s\in S}(w_{qs}^nb_n + \beta_{qs}^n)$ 
    \STATE Compute the allocation probabilities $z_n(\mathbf{b}) = softmax(\bar b_1,\bar b_2,\ldots,\bar b_{N+1};\gamma)$
    \STATE Compute the SPA-0 payments $\theta^0_n(\mathbf{b}) = ReLU(\max_{s\neq n} \bar b_s)$
    \STATE Compute the conditional payment $\theta_l = \Phi_n^{-1}(\theta^0_n(\mathbf{b}))$
    \STATE Compute the loss $\hat{R}(\mathbf{w} ,\bm{\beta})$
    \STATE Update parameters $\mathbf{w}$ and $\bm{\beta}$ using SGD optimizer

    \ENDWHILE
    \RETURN revenue gain by H-AP
  \end{algorithmic}
\end{algorithm}

\section{Experiments}
\subsection{Pre-training of DeepSC}
We utilize the existing open-source code of DeepSC for pre-training \footnote{https://github.com/HQXie0910/DeepSC}. We set the output dimension of encoder $D=d_1=16$, and train the model with AWGN channel for 120 epochs. A total of 7347 English sentences from the proceedings of the European Parliament \cite{koehn2005europarl} are used as the test set for the trained model. Fig. \ref{fig:bleusim} shows that as $D$ decreases, the model has a lower similarity score and 1-gram BLEU score.
\subsection{Evaluation of DL Based Auction}

Following \cite{huang2015performance}, \cite{xu2016wireless}, we set $K=10$, $\eta = 0.8$, $\tau = 1$, $P = 35$ dBm, $E_{cir}=0.5$ mW, $R=2$, $\sigma^2 = -80$ dBm, distance between H-AP and the device $d_{AU} \sim U[8,10]$, and path-loss exponent $\alpha=2$. Entries of $\mathbf{h}$ and $\mathbf{g}$ follow the i.i.d. circularly symmetric complex Gaussian distribution with zero mean and variance $\Omega=10^{-3}(d_{AU})^{-\alpha}$. Following \cite{ju2013throughput}, we assume a 30 dB average signal power attenuation at a reference distance of 1 m in the channel model.

We collected 1000 training samples with $N=10$, $N_s \sim U[15,30]$, $L \sim U[20,32]$, and $b_f=32$. To obtain more realistic training data, we sample the similarity score $s_n \sim U[f_{sim}(D)-\mu_{d},f_{sim}(D)+\mu_{d}]$, where $\mu_d$ is the mean of the difference $f_{sim}(d) - f_{sim}(d-1), \forall d \in [2,d_1]$, and BLEU score is obtained by the same procedure. All experiments are done with two sets of range of $j_n$, i.e., $j_n\sim U[0.1,0.4]$ (higher interest in $BLEU_n$) and $j_n\sim U[0.6,0.9]$ (higher interest in $s_n$). DL based auction is trained with $Q=5$, $S=10$, learning rate of 0.001, and approximate quality $\kappa=1000$. 

\begin{figure}[t]

\begin{minipage}[b]{.48\linewidth}
  \centering
  \resizebox{100pt}{80pt}{%
  \begin{tikzpicture}
\begin{axis}[
    xlabel={Output Dimension $D$},
    xmin=1, xmax=16,
    ymin=0, ymax=0.9,
    legend style={fill=none, nodes={scale=0.8, transform shape}},
    xtick={1,2,3,4,5,6,7,8,9,10,11,12,13,14,15,16},
    ytick={0.1,0.2,0.3,0.4,0.5,0.6,0.7,0.8,0.9,1},
    legend pos=north west,
    ymajorgrids=true,
    xmajorgrids=true,
    grid style=dashed,
    legend cell align={left},
    every axis plot/.append style={very thick},
]

    \addplot[
    color=blue,
    ]
    coordinates {
(1,0.39550235)(2,0.40009948)(3,0.40945041)(4,0.41866887)(5,0.42247792)(6,0.42490115)(7,0.4295931)(8,0.43368545)(9,0.43733177)(10,0.4519554)(11,0.47728359)(12,0.51547686)(13,0.55437698)(14,0.61085957)(15,0.7460733)(16,0.86169747)
    };    \addlegendentry{similarity score};
    
    \addplot[
    color=red,
    dashed,
    ]
    coordinates {
(1,0.0944817)(2,0.09667912)(3,0.09386748)(4,0.10047062)(5,0.10116262)(6,0.10300542)(7,0.11076793)(8,0.11739845)(9,0.12781957)(10,0.15357989)(11,0.1940025)(12,0.27020956)(13,0.34242301)(14,0.44607532)(15,0.65054165)(16,0.82109432)

    }; \addlegendentry{1-gram BLEU score};
\end{axis}
\end{tikzpicture}
  }
  \caption{BLEU and Similarity}
\label{fig:bleusim}
\end{minipage}
\hfill
\begin{minipage}[b]{0.48\linewidth}
  \centering
  \resizebox{100pt}{80pt}{%
\begin{tikzpicture}
\begin{axis}[
    xlabel={No. of Iteration},
    ylabel={Revenue of H-AP},
    xmin=1, xmax=2000,
    ymin=0.75, ymax=0.9,
    xtick={1,500,1000,1500,2000},
    ytick={0.75,0.8,0.85,0.9},
    legend pos=north east,
    legend style={fill=none, nodes={scale=0.8, transform shape}},
    ymajorgrids=true,
    xmajorgrids=true,
    grid style=dashed,
    legend cell align={left},
    every axis plot/.append style={very thick},
    /pgf/number format/.cd,
    1000 sep={},
]
\pgfplotstableread{train_res.txt}\resauction;
\addplot[
    color=red,
    dash pattern=on 6pt off 4pt on 2pt off 4pt,
    ]
    table
    [
    x expr=\thisrowno{0},
    y expr=\thisrowno{1}
    ] {\resauction};
    \addlegendentry{DL-based Auction, $j_n\sim U[0.1,0.4]$};
    
\addplot[
    color=teal,
    ]
    table
    [
    x expr=\thisrowno{0},
    y expr=\thisrowno{2}
    ] {\resauction};
    \addlegendentry{DL-based Auction, $j_n\sim U[0.6,0.9]$};
    
\addplot[
    color=blue,
    dashed,
    dash pattern= on 8pt off 4pt,
    ]
    table
    [
    x expr=\thisrowno{0},
    y expr=\thisrowno{3}
    ] {\resauction};
    \addlegendentry{SPA, $j_n\sim U[0.1,0.4]$};
    
\addplot[
    color=orange,
    dashed,
    ]
    table
    [
    x expr=\thisrowno{0},
    y expr=\thisrowno{4}
    ] {\resauction};
    \addlegendentry{SPA, $j_n\sim U[0.6,0.9]$};
    
\end{axis}
\end{tikzpicture}
  }
  \caption{Revenue of H-AP}
\label{fig:rev}
\end{minipage}
\end{figure}

\begin{figure}[t]

\begin{minipage}[b]{.48\linewidth}
  \centering
  \resizebox{100pt}{80pt}{%
\begin{tikzpicture}
\begin{axis}[
    xlabel={Harvesting Time $\tau$},
    xmin=0.78, xmax=1.22,
    ymin=0.3, ymax=1.2,
    xtick={0.8,0.9,1.0,1.1,1.2},
    ytick={0.3,0.4,0.5,0.6,0.7,0.8,0.9,1.0,1.1,1.2},
    legend pos=north east,
    legend style={fill=none, nodes={scale=0.8, transform shape}},
    ymajorgrids=true,
    xmajorgrids=true,
    grid style=dashed,
    legend cell align={left},
    every axis plot/.append style={very thick},
    /pgf/number format/.cd,
    1000 sep={},
]
\pgfplotstableread{ht_data.txt}\htdata;
\addplot[
    color=red,
    ]
    table
    [
    x expr=\thisrowno{0},
    y expr=\thisrowno{1}
    ] {\htdata};
    \addlegendentry{average highest bid, $j_n \sim U[0.1,0.4]$};
    
\addplot[
    color=red,
    dashed,
    ]
    table
    [
    x expr=\thisrowno{0},
    y expr=\thisrowno{2}
    ] {\htdata};
    \addlegendentry{average bid, $j_n \sim U[0.1,0.4]$};
    
\addplot[
    color=blue,
    mark=*,
    ]
    table
    [
    x expr=\thisrowno{0},
    y expr=\thisrowno{3}
    ] {\htdata};
    \addlegendentry{average highest bid, $j_n \sim U[0.6,0.9]$};

\addplot[
    color=blue,
    mark=*,
    dashed,
    ]
    table
    [
    x expr=\thisrowno{0},
    y expr=\thisrowno{4}
    ] {\htdata};
    \addlegendentry{average bid, $j_n \sim U[0.6,0.9]$};
    
\end{axis}
\end{tikzpicture}
  }
  \caption{Bid vs. Harvest Time $\tau$}
\label{fig:htdata}
\end{minipage}
\hfill
\begin{minipage}[b]{0.48\linewidth}
  \centering
  \resizebox{100pt}{80pt}{%
\begin{tikzpicture}
\begin{axis}[
    xlabel={Distance $d_{AU}$},
    xmin=7.8, xmax=12.2,
    ymin=0.1, ymax=1.2,
    xtick={8,9,10,11,12},
    ytick={0.1,0.2,0.3,0.4,0.5,0.6,0.7,0.8,0.9,1.0,1.1,1.2},
    legend pos=north east,
    legend style={fill=none, nodes={scale=0.8, transform shape}},
    ymajorgrids=true,
    xmajorgrids=true,
    grid style=dashed,
    legend cell align={left},
    every axis plot/.append style={very thick},
    /pgf/number format/.cd,
    1000 sep={},
]
\pgfplotstableread{d_data.txt}\ddata;
\addplot[
    color=red,
    ]
    table
    [
    x expr=\thisrowno{0},
    y expr=\thisrowno{1}
    ] {\ddata};
    \addlegendentry{average highest bid, $j_n \sim U[0.1,0.4]$};
    
\addplot[
    color=red,
    dashed,
    ]
    table
    [
    x expr=\thisrowno{0},
    y expr=\thisrowno{2}
    ] {\ddata};
    \addlegendentry{average bid, $j_n \sim U[0.1,0.4]$};
    
\addplot[
    color=blue,
    mark=*,
    ]
    table
    [
    x expr=\thisrowno{0},
    y expr=\thisrowno{3}
    ] {\ddata};
    \addlegendentry{average highest bid, $j_n \sim U[0.6,0.9]$};

\addplot[
    color=blue,
    mark=*,
    dashed,
    ]
    table
    [
    x expr=\thisrowno{0},
    y expr=\thisrowno{4}
    ] {\ddata};
    \addlegendentry{average bid, $j_n \sim U[0.6,0.9]$};
    
\end{axis}
\end{tikzpicture}
  }
  \caption{Bid vs. Distance $d_{AU}$}
\label{fig:ddata}
\end{minipage}
\end{figure}

\begin{figure}[t]

\begin{minipage}[b]{.48\linewidth}
  \centering
  \resizebox{100pt}{80pt}{%
\begin{tikzpicture}
\begin{axis}[
    xlabel={Sentence Length $L$},
    xmin=19.5, xmax=32.5,
    ymin=0.3, ymax=1.2,
    xtick={20,22,24,26,28,30,32},
    ytick={0.3,0.4,0.5,0.6,0.7,0.8,0.9,1.0,1.1,1.2},
    legend pos=north east,
    legend style={fill=none, nodes={scale=0.8, transform shape}},
    ymajorgrids=true,
    xmajorgrids=true,
    grid style=dashed,
    legend cell align={left},
    every axis plot/.append style={very thick},
    /pgf/number format/.cd,
    1000 sep={},
]
\pgfplotstableread{ml_data.txt}\mldata;
\addplot[
    color=red,
    ]
    table
    [
    x expr=\thisrowno{0},
    y expr=\thisrowno{1}
    ] {\mldata};
    \addlegendentry{average highest bid, $j_n \sim U[0.1,0.4]$};
    
\addplot[
    color=red,
    dashed,
    ]
    table
    [
    x expr=\thisrowno{0},
    y expr=\thisrowno{2}
    ] {\mldata};
    \addlegendentry{average bid, $j_n \sim U[0.1,0.4]$};
    
\addplot[
    color=blue,
    mark=*,
    ]
    table
    [
    x expr=\thisrowno{0},
    y expr=\thisrowno{3}
    ] {\mldata};
    \addlegendentry{average highest bid, $j_n \sim U[0.6,0.9]$};

\addplot[
    color=blue,
    mark=*,
    dashed,
    ]
    table
    [
    x expr=\thisrowno{0},
    y expr=\thisrowno{4}
    ] {\mldata};
    \addlegendentry{average bid, $j_n \sim U[0.6,0.9]$};
    
\end{axis}
\end{tikzpicture}
  }
  \caption{Bid versus Sentence Length $L$}
\label{fig:mldata}
\end{minipage}
\hfill
\begin{minipage}[b]{0.48\linewidth}
  \centering
  \resizebox{100pt}{80pt}{%
\begin{tikzpicture}
\begin{axis}[
    xlabel={Number of Sentences $N_s$},
    xmin=14.5, xmax=30.5,
    ymin=0.25, ymax=1.2,
    xtick={15,18,21,24,27,30},
    ytick={0.2,0.3,0.4,0.5,0.6,0.7,0.8,0.9,1.0,1.1,1.2},
    legend pos=north east,
    legend style={fill=none, nodes={scale=0.8, transform shape}},
    ymajorgrids=true,
    xmajorgrids=true,
    grid style=dashed,
    legend cell align={left},
    every axis plot/.append style={very thick},
    /pgf/number format/.cd,
    1000 sep={},
]
\pgfplotstableread{ns_data.txt}\nsdata;
\addplot[
    color=red,
    ]
    table
    [
    x expr=\thisrowno{0},
    y expr=\thisrowno{1}
    ] {\nsdata};
    \addlegendentry{average highest bid, $j_n \sim U[0.1,0.4]$};
    
\addplot[
    color=red,
    dashed,
    ]
    table
    [
    x expr=\thisrowno{0},
    y expr=\thisrowno{2}
    ] {\nsdata};
    \addlegendentry{average bid, $j_n \sim U[0.1,0.4]$};
    
\addplot[
    color=blue,
    mark=*,
    ]
    table
    [
    x expr=\thisrowno{0},
    y expr=\thisrowno{3}
    ] {\nsdata};
    \addlegendentry{average highest bid, $j_n \sim U[0.6,0.9]$};

\addplot[
    color=blue,
    mark=*,
    dashed,
    ]
    table
    [
    x expr=\thisrowno{0},
    y expr=\thisrowno{4}
    ] {\nsdata};
    \addlegendentry{average bid, $j_n \sim U[0.6,0.9]$};
    
\end{axis}
\end{tikzpicture}
  }
  \caption{Bid versus Number of Sentences $N_s$}
\label{fig:nsdata}
\end{minipage}
\end{figure}

From Fig. \ref{fig:rev}, we observe that the DL based auction constantly achieves higher revenue than that of the SPA, regardless of the performance preference. It is verified that while SPA is IC, it does not maximize the revenue of H-AP. In contrast to SPA, the DL based auction maximizes the revenue of H-AP while keeping the desired properties of IC and IR. Revenue of $j_n \sim U[0.6,0.9]$ is higher because the weight of $s_n$ is higher, and $s_n$ is higher than $BLEU_n$ for every output dimension in our model (Fig. \ref{fig:bleusim}).

Fig. \ref{fig:htdata} shows that increasing wireless energy harvesting time, $\tau$, will increase the average bid of the devices. From Equation \ref{numbit}, the devices have higher data budget when longer harvesting time is given. We also observe that, as harvesting time increases, more devices can achieve their best performance, and hence the average highest bid starts to saturate after some time (1.1s in our settings). From Fig. \ref{fig:ddata}, it is shown that the average bid of devices decreases when they are farther away from H-AP. The reason is that the path loss effect is greater when the distance increases. In our settings, the average highest bid starts to decrease significantly when the distance is more than 9 m. Figs. \ref{fig:mldata} and \ref{fig:nsdata} show that as the size of the text data increases, the average bid decreases regardless of the preference of performance. The reason is that the device had to use fewer features to encode larger data under the energy constraint.

\section{Conclusion}
In this paper, we developed an energy allocation framework for wireless powered semantic communication-based IoT. We derived the valuation of energy based on the critical performance metrics of the semantic communication system, and maximize the revenue of wireless power transmitters while keeping the desired properties of IC and IR using DL based auction. For future work, we can study the energy allocation problem in non-text based transmission such as wireless image and video transmission.

\vfill\pagebreak

\bibliographystyle{IEEEbib}
\bibliography{refs}

\end{document}